\documentclass[12pt,notitlepage, oneside]{report}
\newtheorem{theorem}{Theorem}
\newtheorem{lemma}{Lemma}

\begin{document}
\begin{titlepage}
\title{  \bf \Large Existence of the Solution  \\  
      for the 't Hooft-Polyakov Monopole  } 
\author{     J. B. McLeod and C. B. Wang\thanks{Current address:
  Department of Mathematics, University of Californis, Davis, CA95616. 
  e-mail: cbwang@math.ucdavis.edu. }\\ 
    \it  Department of Mathematics,
        University of Pittsburgh  \\
    \it  Pittsburgh, PA 15260 } 
\date{April 17,  1997}
\maketitle
\begin{abstract}   In this paper we give a mathematical proof of
    the existence of the time independent and
    spherically symmetric solution to the 't Hooft-Polyakov model
    of magnetic monopole by using 2D-shooting method.
\end{abstract}  
\end{titlepage}
\setcounter{equation}{0}
\section*{1. Introduction}
The existence problem of the magnetic monopole has fastinated physicists 
since Dirac's classic work  \cite{dirac} before fifty years ago. In 1974, 
't Hooft \cite{thooft} and Polyakov \cite{polyakov} proposed a model for
 a magnetic monople which arises 
as a static solution of the classical equations for the $SU(2)$ Yang-Mills 
field coupled to an $SU(2)$ Higgs field. The model has been extended by 
Julia and Zee \cite{julia}, and Cho and Maison \cite{cho}. 
In this paper, we discuss the existence and asymptotics
for the solution of 't Hooft-Polyakov monopole.
Prasad and Sommerfield \cite{prasad} found  exact solutions of the field
equation for a special case $(\lambda=0)$.
But so far to our knowledge the boundary value problem for the 
equation of motion with  $\lambda >0$
was just studied numerically. And the existence of the
't Hooft-Polyakov monopole
is just convinced by numerical computations. In this paper
we give
a  mathematical
proof for the existence of the 't Hooft-Polyakov magnetic monopole.

   The Lagrangian of the 't Hooft-Polyakov model \cite{thooft}
\cite{polyakov}   is
    \begin{equation}
    {\cal L}=-{1\over4}F_{\mu \nu}^a F_{\mu \nu}^a
     -{1 \over 2} D_{\mu} \phi_a D_{\mu} \phi_a
     +{1 \over 2} \mu^2 \phi_a \phi_a 
     - {1 \over 4} \lambda ( \phi_a \phi_a)^2,
    \end{equation}
where
    \begin{eqnarray}
    & &F_{\mu \nu}^a = \partial_{\mu} A_{\nu}^a - \partial_{\nu} A_{\mu}^a
      + e \epsilon_{abc}A_{\mu}^b A_{\nu}^c,  \\
    & &D_{\mu} \phi_a = \partial_{\mu} \phi_a
      + e \epsilon_{abc} A_{\mu}^b \phi_c. 
    \end{eqnarray}
The Wu-Yang$^{\cite{wu}}$-'t Hooft-Polyakov Ansatz is 
to seek a solution of the 
equations of
motion (which can be derived from the Lagrangian,see for example \cite{prasad})
 in the form
   \begin{eqnarray}
    & &{A_i}^a = \epsilon_{aij} {{x_j} \over {e r^2}} (1 - K(r)),  \\
    & &{A_0}^a = 0,  \\
    & &\phi_a = {{x_a} \over {e r^2}} H(r).
   \end{eqnarray}
where $K(r), H(r)$ satisfy the equations 
    \begin{eqnarray}
    & & r^2 K''=K(K^2+H^2 - 1),      \label{E:a5}  \\
    & & r^2 H''=2 H K^2 + {\lambda \over e^2} 
          (H^3- e^2 \rho_0^2 r^2 H),        \label{E:a6}
    \end{eqnarray}
where $ \rho_0=\mu / \sqrt{\lambda}$. Let's put
    \begin{equation}
     e= g_0, K(r) = f(r), H(r) = g_0 r \rho(r),
    \end{equation}
then the equations (\ref{E:a5}), (\ref{E:a6}) reduce to 
    \begin{eqnarray}
   & & f^{''} -{{f^2-1} \over {r^2}} f = g_0^2 \rho^2 f,  \label{E:a10} \\
   & & \rho^{''}+{{2} \over {r}} \rho^{'}-{{2} f^2 \over {r^2}}\rho =
      \lambda (\rho^2 -\rho_0^2) \rho,      \label{E:a11}
    \end{eqnarray}
with the boundary conditions  \cite{cho}
    \begin{equation}
    \begin{array}{ll}
     f(0)=1,  &     \rho(0)=0,  \\      
     f(\infty)=0, & \rho(\infty)=\rho_0. 
    \end{array}          \label{E:a12}
    \end{equation}

   Because the exact solution was already found for $\lambda=0$ \cite{prasad},
we just discuss the problem for $\lambda > 0$ in this paper.To solve 
this boundary value problem, we consider
the asymptotics as $r \to 0$,
  \begin{eqnarray}
  & & f(r) \sim 1 - \alpha r^2,    \label{E:a15}  \\
  & & \rho(r) \sim \beta r,           \label{E:a16}
  \end{eqnarray}
where $\alpha, \beta > 0.$ We are going to use
topological 2D-shooting method \cite{mcleod}, \cite{mcleod1}
to prove that there exist such $\alpha, \beta$
such that the corresponding
$f,\rho$ satisfy $f(\infty)=0, \rho(\infty)=
\rho_0$.That is we want to
prove the following theorem in this paper.

\begin{theorem} For $\lambda >0, \rho_0>0, g_0 > 0$, there is a 
solution $(f,\rho)$ to equations
(\ref{E:a10}), (\ref{E:a11}) satisfying
the boundary conditions (\ref{E:a12}), and
  \begin{eqnarray}
  0<f<1,  &  0<\rho< \rho_0,  \\
  f^{'} <0,  &  \rho^{'} > 0,
  \end{eqnarray}
for $0<r<\infty$.
\end{theorem}

   The plan of this paper is as follows. In sec.2,
we consider the
asymptotic behaviour of the solution to the eqs.
(\ref{E:a10}), (\ref{E:a11}).Then we discuss the behaviours
of $f(r)$ as $\alpha$ small or large for $\beta$ in a
finite interval. We show that $f'$ cross $0$
when $\alpha$ small in sec. 3, and $f$ cross $0$ when
$\alpha$ large in sec. 4. In sec. 5 we talk about the
possibilities of the behaviours for $\rho$ when $f$ stays between $0$
and $1$. While $f$ stays between $0$ and $1$, we  show
that when $\beta$ is small $\rho$
crosses $0$  in sec. 6, and
as $\beta$ large $\rho$ crosses $\rho_0$ in sec. 7.  Finally
by the topological lemma (McLeod and Serrin \cite{mcleod})
we show that the solution to the boundary value problem
(\ref{E:a10}), (\ref{E:a11}) and (\ref{E:a12}) exists.

\setcounter{equation}{0}
\section*{2. Asymptotics of Solution at the Origin}

\begin{lemma}For any $\alpha, \beta$, there
is a unique solution
$(f,\rho)$ to the equations (\ref{E:a10}),(\ref{E:a11}) , such that
  \begin{eqnarray}
  f & \sim& 1 - \alpha r^2,   \\
  \rho &\sim& \beta r,
  \end{eqnarray}
as $ r \to 0$.
\end{lemma}
{\it Proof \,} Set
  \begin{equation}
  s=\log r, f(r)=1 + p(s), \rho(r) = q(s).
  \end{equation}
Then eqs. (\ref{E:a10}),(\ref{E:a11})    are reduced to
  \begin{eqnarray}
  & &  p^{''} - p^{'}-2 p=3 p^2+p^3+g_0^2 e^{2s} q^2 p,  \\
  & & q^{''} + q^{'}-2q=
      2(2p+p^2)q + \lambda e^{2s}(q^2-\rho_0^2)q,
  \end{eqnarray}
which is at least formally equivalent to the
integral equations
  \begin{eqnarray}
   p(s)=-\alpha e^{2s}+{1\over3}\int_{-\infty}^s \,
     (e^{2(s-\sigma)}-e^{-(s-\sigma)} )
     (3p^2+p^3+g_0^2 e^{2\sigma}q^2 p) \, d\sigma, \label{E:b10} \\
   q(s)=\beta e^s +{1\over2} \int_{-\infty}^s \,
    (e^{s-\sigma}-e^{-2(s-\sigma)} )
   (2 (2p+p^2)q+\lambda e^{2\sigma}(q^2-\rho^2)q)
           \, d\sigma.   \label{E:b11}
   \end{eqnarray}

   Let
 $$  p=e^{2s} \phi, q= e^s \psi.  $$
The eqs.  (\ref{E:b10}) and (\ref{E:b11})  become
 \begin{eqnarray}
 \phi =-\alpha+{1\over3}\int_{-\infty}^s \,
    (e^{2 \sigma}-e^{-3s+5 \sigma} )
    (3 \phi^2+e^{2 \sigma} \phi^3+
      g_0^2 e^{2\sigma}\psi^2 \phi) \, d\sigma, \\
 \psi =\beta +{1\over2}\int_{-\infty}^s \,
    (e^{2 \sigma}-e^{-3s+5\sigma})
    (2(2\phi+e^{2\sigma}\phi^2)\psi+
    \lambda(e^{2\sigma}\psi^2-\rho_0^2)\psi) \, d\sigma ,
 \end{eqnarray}
which can be solved by iteration by setting
   \begin{eqnarray*}
  & & \phi_0 = - \alpha,  \psi_0=\beta, \\ 
  & & \phi_{n+1}(s) = -\alpha+{1\over3}\int_{-\infty}^s \,
    (e^{2 \sigma}-e^{-3s+5 \sigma} )
    (3 \phi_n^2+e^{2 \sigma} \phi_n^3+
      g_0^2 e^{2\sigma}\psi_n^2 \phi_n) \, d\sigma, \\
  & & \psi_{n+1}(s) = \beta +{1\over2}\int_{-\infty}^s \,
    (e^{2 \sigma}-e^{-3s+5\sigma})
    (2(2\phi_n+e^{2\sigma}\phi_n^2)\psi_n+
    \lambda(e^{2\sigma}\psi_n^2-\rho_0^2)\psi_n) \, d\sigma,
   \end{eqnarray*}
for $n\ge 0$.

   We choose constant numbers
$K, S$ by
  \begin{eqnarray}
  K &=& \max(2 |\alpha|, 2 |\beta|, 3),  \\
  e^{2S} &=& \max(M_1,M_2,M_3,M_4),
  \end{eqnarray}
where
   \begin{eqnarray*}
 M_1 &=& {1 \over 2} (2(2 K+K^2)+\lambda(K^2+\rho_0^2) ),\\ 
 M_2 &=& {1 \over 3}(6 K+2 K^2+2 g_0^2 K^2),  \\
 M_3 &=& (8 + 3 \lambda) K^2 + \lambda \rho_0^2,  \\
 M_4 &=& 4 K + 3(1 + g_0^2) K^2.
   \end{eqnarray*} 
We claim that for $s \in (-\infty, -S]$, there are
   \begin{eqnarray}
   | \phi_n |, | \psi_n | &\le& K,  \\
   | \phi_{n+1}-\phi_n | &\le& {M \over 3^{n+1}} e^{2s},\\
   | \psi_{n+1}-\psi_n | &\le& {M \over 3^{n+1}} e^{2s},
   \end{eqnarray}
for $n \ge 0$, where
    \begin{eqnarray}
    M=max(4 K^3+\lambda(K^2+\rho_0^2)K,(2+g_0^2)K^3).
    \end{eqnarray}
This can be proved by induction. We skip the details here.
Hence $\{ (\phi_n,\psi_n) \}_{n=0}^\infty $ is convergent.
The uniqueness is similar.

\setcounter{equation}{0}
\section*{3. The Solutions for Small $\alpha$ }

\begin{lemma} For any $\beta_2 \ge \beta_1 >0$, when
$\beta \in [\beta_1, \beta_2]$, there exists $\alpha_1>0$, such that
if $\alpha \in (0, \alpha_1]$, there is $r^- >0$, so that
  \begin{eqnarray}
  f^{'}(r^-) &>& 0,   \\
  f(r) &>& 0, \textnormal{ for \,} 0<r \le r^-.
  \end{eqnarray}
\end{lemma}
{\it Proof \,} We make a scaling
   \begin{equation}
     t={r \over \sqrt{\alpha} }, \, f(r)=1+\alpha^2 p(t), \,
     \rho(r) = \sqrt{\alpha} q(t).
   \end{equation}
Then the equations become
  \begin{eqnarray}
    p^{''}- {2p+\alpha^2 p^2  \over t^2}(1+\alpha^2 p)=
          g_0^2 q^2(1 + \alpha^2 p), \label{E:c5}  \\
    q^{''}+{2 \over t} q^{'}-{2(1+\alpha^2 p)^2 \over t^2} q =
         \lambda \alpha(\alpha q^2- \rho_0^2) q, \label{E:c6}
  \end{eqnarray}
 with the asymptotics at the origin
   \begin{eqnarray}
     p(t) &\sim& - t^2,   \label{E:c7}   \\
     q(t) &\sim& \beta t,  \label{E:c8}
   \end{eqnarray}
 as $t \to 0$.

    Letting $\alpha =0 $, the problem is reduced to
    \begin{eqnarray}
   & &  P(t) \sim -t^2, Q(t) \sim \beta t,  t \to 0,   \\
   & &  P^{''}- {2 P \over t^2} = g_0^2 Q^2,    \\
   & &  Q^{''}+ {2 \over t} Q^{'} - {2 \over t^2} Q =0.
    \end{eqnarray}
 There is a unique solution for this problem. It's not difficult
 to see that the solution is
   \begin{eqnarray}
      P(t)&=&- t^2 + {g_0^2 \beta^2 \over 10} t^4,   \\
      Q(t) &=& \beta t,
   \end{eqnarray}
 and then
    \begin{equation}
    P^{'}(t)=2t(-1+{g_0^2 \beta^2 \over 5} t^2).
    \end{equation}

   For $\beta \in [\beta_1, \beta_2]$, choose small $\epsilon^->0$,
and
   \begin{equation}
   t_0 ={\sqrt{5} \over g_0 \beta_1} +\epsilon^-,
   \end{equation}
such that
   \begin{eqnarray}
   P^{'}(t_0) &>&0,  \\
   | P(t) | &\le& {t_0}^2 +{g_0^2 \beta_2^2 \over 10} {t_0}^4,
                  0 \le t \le t_0.
   \end{eqnarray}
Now for the solution $p,q$ to the equations (\ref{E:c5}), (\ref{E:c6})
  with the conditions (\ref{E:c7}) and (\ref{E:c8}),
since $p,p^{'},q$ are contineous in $r, \alpha, \beta$, by uniform
continuity in compact set, there exists $\alpha_1>0$ satisfying
  $$ 2 \alpha_1^2 < (t_0^2+{{g_0^2 \beta_2^2} \over 10} t_0^4)^{-1},  $$
such that if $\alpha \in (0, \alpha_1], \beta\in[\beta_1,\beta_2]$,
     \begin{eqnarray}
     p^{'}(t_0) &>& 0,   \\
	  | p(t) | &\le&
		2 \left( {t_0}^2 +{g_0^2 \beta_2^2 \over 10} {t_0}^4 \right),
							 0 \le t \le t_0,
     \end{eqnarray}
which implies that for $\alpha \in (0, \alpha_1] $
  \begin{eqnarray*}
   f^{'}(r^-) &=& \alpha^{3/2}p^{'}({r^- \over \sqrt{\alpha}}) >0,  \\
   f(r) &>& 1- \alpha^2 | p(t) |  \\
     &\ge& 1- 2 \alpha_1^2 
       \left(  {t_0}^2 +{g_0^2 \beta_2^2 \over 10} {t_0}^4 \right) >0,
                   \, \,    0<r \le r^{-},
   \end{eqnarray*}  
where $r^- = \sqrt{\alpha} t_0$. So the lemma is proved.

\setcounter{equation}{0}
\section*{4. The Solutions for Large $\alpha$}

\begin{lemma} For any $\beta_2 \ge \beta_1 \ge 0$, when
$\beta \in [\beta_1, \beta_2]$, there is  a large $\alpha_2>0$, such that
if $\alpha \in [\alpha_2,\infty)$, there exists $r^+ >0$, so that
  \begin{eqnarray}
  f(r^+) &<& 0,   \\  \label{E:d5}
  f^{'}(r) &<& 0,  0<r \le r^+.   \label{E:d6}
  \end{eqnarray}
\end{lemma}
{\it Proof \,} We make another scaling
    \begin{equation}
    t=\sqrt{\alpha} r, f(r)=1-\psi(t), \rho(r) = \phi(t).
    \end{equation}
Then the equations become
   \begin{eqnarray}
   \psi^{''}- {\psi(1-\psi)(2-\psi) \over t^2}
            ={-1 \over \alpha}g_0^2 \phi^2(1-\psi),  \\
   \phi^{''}+{2 \over t} \phi{'}-{2(1-\psi)^2 \over t^2} \phi
            ={1 \over \alpha}\lambda(\phi^2-\rho_0^2))\phi,
   \end{eqnarray}
with the asymptotics as $t \to 0$
   \begin{eqnarray*}
   \psi(t) &\sim& t^2,  \\
   \phi(t) &\sim& {1 \over \sqrt{\alpha} } \beta t.
   \end{eqnarray*}

   Then as $\alpha \to \infty$, $\phi \to 0$ uniformlly on compact
intervals in $t$, while $\psi$  tends, also uniformlly on compact
intervls in $t$, to the solution $\Psi$ of
   \begin{eqnarray*}
  & &  \Psi^{''} = { \Psi(1-\Psi)(2-\Psi) \over t^2},   \\
  & &  \Psi(t) \sim t^2, t \to 0.
   \end{eqnarray*}
To get the behaviour of $\Psi$, we make a transformation
 $$  s = \log t.  $$
Then the equation is reduced to
   \begin{eqnarray*}
  & & \Psi_{ss}= \Psi_s+2\Psi- 3 \Psi^2+\Psi^3,  -\infty <s <\infty, \\
  & & \Psi(s) \sim e^{2s},  s \to -\infty.
   \end{eqnarray*}
Multiplying this equation by$d\Psi/ds$ and integrating,
we arrive at
  $$  {1 \over 2} \Psi_s^2 = \Psi^2(1-{\Psi \over 2} )^2 +
               \int_{-\infty}^s\, \Psi_\sigma^2 \, d\sigma.   $$
This makes clear that $d\Psi/ds$ does not vanish and $\Psi$ becomes
unbounded and certainly crosses 1, while $d \Psi/ds$ keeps positive
at least before the crossing.

   By the same argument as in lemma 1, we see that for
$\beta \in[\beta_1, \beta_2]$, there is $\alpha_2 >0$, such that
if $\alpha \ge \alpha_2$, there exists $r^+ =r^+(\alpha)$,
so that   the lemma     holds.

\setcounter{equation}{0}
\section*{5. Argument for 
$\alpha \notin S_{\beta} ^{-} \cup S_{\beta}^{+}$ }
   
   For any $\beta > 0$, define
   \begin{eqnarray*}
  & & S_{\beta}^{-}=\{ \alpha >0 | 
    f' \, \textnormal{ cross 0 before} \,f\, \textnormal{ cross} 0 \}, \\ 
  & & S_{\beta}^{+}=\{ \alpha >0 | 
     f \, \textnormal{ cross 0 before} \,f'\, \textnormal{ cross} 0 \}.
   \end{eqnarray*}
By Lemma 2,3, $S_{\beta}^{-}$ and $S_{\beta}^{+}$ are not empty and
disjoint.  
By implicit function theorem, it's not hard to prove that
$S_{\beta}^{-}, S_{\beta}^{+}$ are open sets, so that
$(0, \infty) \setminus  (S_{\beta}^{-} \cup S_{\beta}^{+})$ is  not empty
and closed set. For 
$\alpha \in (0, \infty) \setminus (S_{\beta}^{-} \cup S_{\beta}^{+})$,
we simply denote it as 
$\alpha \notin (S_{\beta}^{-} \cup S_{\beta}^{+})$. By 
eq. (\ref{E:a10})  we see that
if $f=0,f'=0$ at the same time, then the $f$ is identically zero, which 
is impossible. So we have proved the following lemma.
\begin{lemma} If $\alpha \notin (S_{\beta}^{-} \cup S_{\beta}^{+})$, then
     $$ 0<f<1, f' \le 0, $$
for $0<r<\infty$.
\end{lemma}
\begin{lemma} For $\beta >0, \alpha \notin S_{\beta}^{-} \cup S_{\beta}^{+}$,
there are three possibilities for $\rho$,
  
	(A) $\rho' $ cross $0$ at some point $r=r_0$, while
		 $0<\rho<\rho_0$, for $0<r \le r_0$.

	(B) $\rho$ cross $\rho_0$.

	(C) $0<\rho<\rho_0, \rho' \ge 0$, for $0<r<\infty$ ,and
		\begin{eqnarray}
       & &  \rho(\infty) =\rho_0,  \\
		 & &  f(\infty) =0.
      \end{eqnarray}
\end{lemma}
{\it Proof \,}
 Because $\rho'(0) = \beta >0$, if  $\rho$ does not cross $\rho_0$
(case (B)) ,
then either  $\rho' $ crosses $0$ at some point $r=r_0$, while
$0<\rho<\rho_0$, for $0<r \le r_0$(case(A)) , or
$\rho' \ge 0, 0<\rho<\rho_0$,  for $0<r<\infty$(case(C)) .
There is no possibility that when  $\rho$ does not cross $\rho_0$,
 $\rho=\rho_0$  at some point.
In fact,
if $\rho=\rho_0$ and $\rho^{'} =0$ at the same time, then by eq. (\ref{E:a11})
we have $\rho^{''} > 0$ at this point, since
$\beta >0, \alpha \notin S_{\beta}^{-} \cup S_{\beta}^{+}$. Then
 $\rho$  crosses $\rho_0$, which is a contradiction.

  For case (C), we have
	\begin{eqnarray}
f(0)=1,0<f <1,\, f^{'} \le 0,\,
		  f^{'}(\infty) = 0,\, f(\infty)=a, \label{E:e5} \\
\rho(0)=0, 0<\rho<\rho_0,\, \rho^{'} \ge 0,\, \rho^{'}(\infty)=0, \,
		  \rho(\infty) =b,  \label{E:e6}
	 \end{eqnarray}
for some $a \in [0,1), b \in (0,\rho_0]$, where the second and third parts
of (\ref{E:e5}), (\ref{E:e6}) are for $ 0 < r < \infty$. We want to show
$a=0, b = \rho_0$.

   Suppose $b < \rho_0$. Choose $r_1>0$, so that
$ \, b/2 \le \rho \, $, for $\, r_1 \le r <\infty \,$.   
By eq. (\ref{E:a11}), there is
   \begin{eqnarray*}
 (r^{2} \rho^{'})^{'} &=& 
        \rho+\lambda r^2 \left( \rho^2-\rho_0^2 \right) \rho \\
              &\le& 
      b - \lambda r^2 \left( \rho_0^2-b^2 \right){b\over 2},
   \end{eqnarray*}  
for $r_1 \le r <\infty$. Integrting from $r_1$ to $r$,
and dividing  $r^2$ both sides, finally we get
   \begin{eqnarray*}
	  \rho^{'}(r) & \le& {r_1^2 \over r^2} \rho^{'}(r_1) +
        b \left( {1 \over r}-{r_1 \over r^2} \right) -
        {\lambda b \over 6}
       \left( \rho_0^2-b^2 \right) \left( r-{r_1^3 \over r^3} \right) \\
      &\to & - \infty, \textnormal{ as \,} r \to \infty.
   \end{eqnarray*}  
This is a contradiction. So $b=\rho_0$.

    Now suppose $a > 0$. Choose $r_2 >0$ so that
    \begin{eqnarray*}
    {1 - a^2 \over r^2} &\le& {g_0^2 \rho_0^2 \over 8}, \\
    \rho &\ge& \rho_0/2,
    \end{eqnarray*}
for $r_2 \le r <\infty$.  Then we have from eq. (\ref{E:a10})
   \begin{eqnarray*}
  f^{''} &=& f({{f^2-1} \over {r^2}}  + g_0^2 \rho^2 ) \\
         &\ge& f ({{a^2 -1} \over {r^2}} +
                  {{g_0^2 \rho_0^2} \over {4}} )  \\
         &\ge& f {{g_0^2 \rho_0^2} \over {8}}   \\
         &\ge& {{a g_0^2 \rho_0^2} \over {8}}.
  \end{eqnarray*}  
Integrating from $r_2$ to $r$, we get
 $$  f^{'}(r) \ge 
        f^{'}(r_2)+ {a g_0^2 \rho_0^2 \over 8}\left( r-r_2 \right)
          \to \infty, \textnormal{ as \,} r \to \infty.  $$
This is a contradictin. So $a=0$. So the lemma is proved. 

\setcounter{equation}{0}
\section*{6. The Solutions for Small $\beta$}
   
  We make a transformation
    \begin{equation}
     t=\sqrt{\lambda} \rho_0 r,f(r)=p(t), 
        \rho(r) = \beta q(t).
    \end{equation}
Then the equation and the asymptotics become
   \begin{eqnarray}
 &  & p^{''} -{p^2-1 \over t^2} p =
        \beta^2 g_{1}^2  q^2 p,   \label{E:f5}  \\
 &(P)&  q^{''}+{2 \over t} q^{'}+
     (1-{2 p^2 \over t^2} ) q =
      {\beta^2 \over \rho_0^2} q^3,    \label{E:f6}  \\
 &  & p(t) \sim 1 - {\alpha \over \lambda \rho_0^2} t^2,  
     q(t) \sim {t \over \sqrt{\lambda} \rho_0 }, \,
       \textnormal{as $t \to 0$},  \label{E:f7}
   \end{eqnarray}
where $g_1 =\lambda \rho_0^2$, and we will consider 
$\alpha \ge 0, \beta \ge 0$.
The difference between this system and the system 
(\ref{E:a10}), (\ref{E:a11}) and (\ref{E:a12}) is only for the case 
$\beta =0$,
because when $\beta \ne 0$ these two systems are equivalent. But for
$\beta =0$ we can show by the same method that this problem has
unique solution. For $\beta=0, \alpha>0$, by the same argument
as in the proof of Lemma 3, we see that $p$ cross $0$. And for
$\beta=0, \alpha=0$, there is $p=1$.
 
  Let's choose $\bar{\beta} >0$. By Lemma 3 for
$\beta \in [0, \bar{\beta}]$ there is 
$\bar{\alpha} >0$, such that for any $\alpha \ge \bar{\alpha}$, we have
$\alpha \in S_{\beta}^{+}$. Now let's define in the $(\alpha, \beta)$ 
plane the sets
  \begin{eqnarray}
  & & D = [0,\bar{\alpha}] \times [0,\bar{\beta}],  \\
  & & D^{-} = \{(\alpha,\beta) \in D | 
     p' \, \textnormal{cross 0 before \, $p$ \, cross 0} \}, \label{E:ff5} \\
  & & D^{+} = \{(\alpha,\beta) \in D | 
     p \, \textnormal{cross 0 before \, $p'$\, cross 0} \}, \label{E:ff6} \\
  & & l = \{0\} \times (0, \bar{\beta} ],   \\
  & & D_{0} = D \setminus (l \cup D^{-} \cup D^{+}).
  \end{eqnarray}
We have that $D^{-}, D^{+}$ are open in $D$, not empty and disjoint.
By Lemma 2, we see that $l \cup D^{-}$ is open in $D$, so 
 $D_{0}$ is compact. Note that for any $\beta \in [0, \bar{\beta} ]$,
there is $D_0 \cap ([0,\bar{\alpha}] \times \{\beta\}) \neq \emptyset$.
And  for $\beta=0$, only $(0,0) \in D_0$.

    Also let's set
     \begin{equation}
    \mathbf{B}=\{ p | \textnormal{$(p,q)$ is a solution to (P)
      for $(\alpha, \beta) \in D_0$ }  \}.
     \end{equation}
We see that for any $p \in \mathbf{B}$, the properties in (\ref{E:ff5}),
(\ref{E:ff6}) are not satisfied, and $p, p'$ can not vanish at the same time
by uniqueness of the solution. So we get
$0 < p(t) \le 1$, for $0<t<\infty$, if $p \in \mathbf{B}$.

    Now let's restrict $(\alpha,\beta)$ in $D_0$.
    
    We have already seen that in $D_0$, the problem (P)  has a unique
solution $(p,q)$ satisfying $0<p \le 1$, and by eq. (\ref{E:f5}) 
   we see that
$q$ is bounded in any finite interval for $\beta >0$.And for
$\beta=0, \alpha =0$, there are $p=1,q$ is expressed by
$J_{3/2}(t)$ ( see (\ref{E:f22}) for $p=1$). Thus we get the following
result.
     
\begin{lemma} For $(\alpha, \beta) \in D_0$, the problem (P)
   has a unique solution
and for any finite $\bar{t} >0$, $q(t)$ is uniformly bounded 
for
$(t, \alpha,\beta) \in [0,\bar{t}] \times D_0$, and 
$0<p \le 1$ for $(t,\alpha, \beta) \in [0, \infty) \times D_0$.
\end{lemma}

	Next, we want to find two linearly independent solutions of
the equation
  \begin{equation} 
   Q''+{2 \over t} Q'+(1-{2 p^2 \over t^2} )Q =0.  \label{E:f20}
  \end{equation}
Let
  $$    Q(t) = {1 \over {\sqrt{t}} } y(t),  $$
which reduces the equation (\ref{E:f20}) to
   \begin{equation}
  y''+{1 \over t} y' + \left( 1-{\nu^{2} \over t^{2} }+
            {2(1-p^{2}(t)) \over t^{2} } \right) y =0, \label{E:f22}
   \end{equation}
where $\nu =3/2$. 

  Consider the Bessel functions 
  \begin{eqnarray}   
  & & J_{3/2}(t) = \left( {2 \over {\pi t}} \right)^{1/2}
    \left( { \sin{t} \over t} - \cos{t} \right),  \\
  & & J_{-3/2}(t) = - \left( { 2 \over {\pi t} } \right)^{1 /2}
    \left( {\cos{t} \over t} + \sin{t} \right).
  \end{eqnarray} 
Let $t_0$ be the first positive zero of $J_{3/2}(t)$.
 Choose $t_1>t_0$, so that $J_{3/2}(0) =J_{3/2}(t_0)=0,
J_{3/2}(t) >0$, for $t \in (0,t_0)$, and $J_{3/2}(t) <0$,
for $t \in (t_0,t_1)$.

\begin{lemma} For each $(\alpha,\beta) \in D_0, p \in \mathbf{B}$, 
there are two
linearly independent solutions $y_p^{(1)}(t), y_p^{(2)}(t)$ 
to the equation (\ref{E:f22})
uniquely determined by the asymptotics
 \begin{eqnarray}
 & &  y_p^{(1)}(t) \sim J_{3/2}(t) 
    \sim {1 \over 3} \left( {2 \over \pi} \right)^{1/2} t^{3/2} ,  \\
 & &  y_p^{(2)}(t) \sim J_{-3/2}(c_0 t)
    \sim - \left( {2 \over \pi} \right)^{1/2} \left( c_0 t \right)^{-3/2}, 
  \end{eqnarray}
as $t \to 0$, where 
   $$  
   c_0=\sqrt{1+{{4 \alpha} \over {\lambda \rho_0^2}}} . 
   $$
There is singularity only for $y_p^{(2)}$ at the origin.
\end{lemma}
{\it Proof \,} First of all we have the Wronskian
   $$ W \left( J_{3/2}, J_{-3/2} \right) = {2 \over {\pi t}}. $$
Define $y_p^{(1)}$ for $0<t<\infty$ by the integral equation
  \begin{eqnarray*} 
   y_p^{(1)}(t) = J_{3/2}(t) &+&
     \pi J_{-3/2}(t) \int_0^t \, \sigma J_{3/2}(\sigma) 
      { { p^2(\sigma)-1 } \over \sigma^2 } y_p^{(1)}(\sigma) \, d\sigma \\
  & -& \pi J_{3/2}(t) \int_0^t \, \sigma J_{-3/2}(\sigma)
      { { p^2(\sigma)-1 } \over \sigma^2 } y_p^{(1)}(\sigma) \, d\sigma.
   \end{eqnarray*}
By iteration method one can show that $y_p^{(1)}(t)$ is uniquely defined
without singularity.

   Now let's consider how to define  $y_p^{(2)}(t)$.By eq. (6.2) it's not 
hard to see that 
  $$p(t) = 1 - { \alpha \over \lambda \rho_0^2} t^2 + {\large O}(t^4), $$
as $t \to 0$. Let
  \begin{eqnarray*}
    & & s= c_0 t,   \\
    & & y(t) = z(s)  \\
    & &R(s) =c_0^{-2} \left( { 4 \alpha \over \lambda \rho_0^2}-
            {2(1-p^2(t) \over t^2}
             \right).
  \end{eqnarray*} 
By simple calculation we see that 
    $$ R(s) = {\large O}(s^2),  $$
as $s \to 0$.
So now eq. (6.11) becomes
    $$ z'' + {1 \over s} z'+\left(1-{\nu^2 \over s^2}- R(s) \right) z = 0.
    $$
Define $z(s)$ by
  \begin{eqnarray*}
 &   z(s) = J_{-3/2}(s) &+
     {\pi \over 2} J_{-3/2}(s) \int_0^s \, \sigma J_{3/2}(\sigma) 
      R(\sigma) z(\sigma) \, d\sigma  \\
 &  &  - {\pi \over 2} J_{3/2}(s) \int_0^s \, \sigma J_{-3/2}(\sigma)
      R(\sigma) z(\sigma) \, d\sigma, \\
 &  z(s) \sim J_{-3/2}(s), & \textnormal{as $s \to 0$}.
   \end{eqnarray*}
Now let $y_p^{(2)}(t)=z(s)$, and then the lemma is done.

  So now
   \begin{equation}
      Q_p^{(j)}(t) = {1 \over \sqrt{t} } y_p^{(j)}(t), j=1,2
   \end{equation}  
forms a basis of eq. (\ref{E:f20}) with the Wronskian
  \begin{equation}
   W(Q_p^{(1)},Q_p^{(2)}) = { \gamma \over t^2},
  \end{equation}
where 
 $$\gamma = \lim_{t \to 0} t^2 W(Q_p^{(1)},Q_p^{(2)}) = 
        {5 \over 3 \pi} c_0^{-3/2}. $$  

\begin{lemma}
   \begin{equation}
     m=\sup_{p \in \mathbf{B} } \inf_{t \in [0,t_1]} y_p^{(1)}(t) <0,
  \end{equation}
and
   \begin{equation}
       m_0=\sup_{p \in \mathbf{B} } \inf_{t \in [0,t_1]} Q_p^{(1)}(t) <0.
  \end{equation}
\end{lemma}
{\it Proof \,} Because $J_{3/2}(0) =0=J_{3/2}(t_0)$, and for any
$p \in \mathbf{B}, 0<p \le 1$, by Sturm comparison principle, there is
a zero of $ y_p^{(1)}(t)$ between $0$ and $t_0$. Since  $ y_p^{(1)}(t)$
is not a trival solution, by the uniqueness theorem, $ (y_p^{(1)}(t))'$,
$ (y_p^{(1)}(t))''$ can be zero at the same time. Thus  $ y_p^{(1)}(t)$
must cross 0 in $(0, t_1)$ for any $p \in \mathbf{B}$, which implies
   \begin{equation}
    \inf_{t \in [0,t_1]} y_p^{(1)}(t) <0. \label{E:f31}
   \end{equation}
So $m \le 0$.

   Suppose $m=0$. Then there is a sequence 
$\{ (\alpha^{(n)}, \beta^{(n)} ) \}_{n=1}^{\infty} \subset D_0$, such that
  $$ \inf_{t \in [0,t_1]} y_{p^{(n)}}^{(1)}(t) \to 0,  $$
as $ n \to \infty$, where $ p^{(n)} = p(t, \alpha^{(n)}, \beta^{(n)} )$.
Because $D_0$ is compact, without loss of generility, assume
  $$ (\alpha^{(n)}, \beta^{(n)} ) \to (\alpha^*, \beta^*) \in D_0, $$
as $ n \to \infty$.
By continuity, there is
  $$ \inf_{t \in [0,t_1]} y_{p^*}^{(1)}(t) = 0, \, \,
    p^* = p(t, \alpha^*, \beta^* ) $$ 
This is a contradiction because $(\alpha^*, \beta^*) \in D_0$, which 
implies $p^*$ satisfies (\ref{E:f31}). Thus $m<0$. And if $m_0 \ge 0$, 
then $m \ge 0$.
So we also have $m_0 < 0$.

\begin{lemma} There is a small $\beta_1 > 0$, such that for any
$\beta \in (0, \beta_1]$, 
if $\alpha \notin S_\beta^{-} \cup S_\beta^{+}$, then (A) is satisfied.
\end{lemma}
{\it Proof} Suppose $(p,q)$ is a solution to 
(\ref{E:f5}), (\ref{E:f6}) and (\ref{E:f7}), then by the variation
of parameter, $q$ satisfies the integral equation
    \begin{equation}
  q(t) = \mu Q_p^{(1)}(t) + 
      {\beta^2 \over {\gamma  \rho_0^2}} G(t), 0 \le t \le t_1, \label{E:f40}
   \end{equation}
where
   \begin{equation}
    G(t) =Q_p^{(2)}(t) \int_0^t \, s^2 Q_p^{(1)}(s) q(s)^3 \, ds -
         Q_p^{(1)}(t) \int_0^t \, s^2 Q_p^{(2)}(s) q(s)^3 \, ds,
   \end{equation} 
and
  $$ \mu =  {3 \over \rho_0} \sqrt{ {\pi \over {2 \lambda}} } .$$

   By Lemma 6,7, we see that  $G(t)={\large O}(t^3)$, as $t \to 0$, 
and $G$ is uniformly bounded in $[0,t_1] \times D_0$.
Choose $\bar{\beta_1} < \bar{\beta}$, so that if $0<\beta<\bar{\beta_1}$,
there is
  $$ \left| {\beta^2 \over {\gamma  \rho_0^2}}  G(t) \right| < 
     {{\mu |m_0|} \over 2},
  $$
which implies by (\ref{E:f40}) and by Lemma 8
  $$\inf_{t \in [0,t_1]} q(t) < 
 \inf_{t \in[0, t_1]} \left(\mu Q_p^{(1)}(t) + {{\mu |m_0|} \over 2} \right) <
     {{\mu m_0} \over 2}  <0,
  $$ 
which means $q$ crosses $0$, since $q'(0) = 1$.
And then $q'$ or $\rho'$ cross 0.
Choose smaller positive $\beta_1 < \bar{\beta_1}$, such that $0<\rho< \rho_0$
before $\rho'$ crosses $0$.So 
the lemma is proved.

\setcounter{equation}{0}
\section*{7. The Solutions for Large $\beta$}

\begin{lemma} There is a large $\beta_2 > 0$, such that
if $\alpha \notin S_\beta^{-} \cup S_\beta^{+}$, then (B) is satisfied.
\end{lemma}
{\it Proof \,} Recall the integral equation form we used in
section 2,
 \begin{eqnarray}
   \phi=-\alpha+{1\over3}\int_{-\infty}^s \,
    (e^{2 \sigma}-e^{-3s+5 \sigma} )
    (3 \phi^2+e^{2 \sigma} \phi^3+
      g_0^2 e^{2\sigma}\psi^2 \phi) \, d\sigma, \label{E:g10} \\
   \psi=\beta +{1\over2}\int_{-\infty}^s \,
    (e^{2 \sigma}-e^{-3s+5\sigma})
    (2(2\phi+e^{2\sigma}\phi^2)\psi+
    \lambda(e^{2\sigma}\psi^2-\rho_0^2)\psi) \, d\sigma, \label{E:g11}
 \end{eqnarray}
where
  \begin{equation}
  s=\log r, f(r)=1 + e^{2s} \phi(s), \rho(r) = e^{s} \psi(s).
  \end{equation}

    Suppose there is a sequence 
$\{ (\alpha^{(n)}, \beta^{(n)}) \}_{n=1}^{\infty} \subset 
(0, \infty) \times (0, \infty) $ with 
$\beta^{(n)} \to \infty$, 
as $n \to \infty$, and  
$\alpha^{(n)} \notin S_\beta^{-} \cup S_\beta^{+}$,
 such that (B) is not satisfied for each $n$. By
Lemma 5   for each n, either (A) or (C) is satisfied. If (C) is satisfied,
the theorem is proved. So we consider for each $n$, (A) is satisfied. 
Let us denote
   \begin{eqnarray*}
     & &   f^{(n)} = f(r, \alpha^{(n)}, \beta^{(n)} ),  \\
    & & \rho^{(n)} = \rho(r, \alpha^{(n)}, \beta^{(n)} ).
   \end{eqnarray*}
   For each $n$, let $[0, r_n]$ be the maximal interval such that
    \begin{equation}
      | \rho^{(n)} | \le \rho_0.     \label{E:g50}
    \end{equation}
Let
   $$    \bar{r} = \inf_{n} (r_n). $$
We want to show $\bar{r} > 0$(maybe $\infty$). If $\bar{r} =0$, without 
loss of generality, assume $r_n \to 0$, as $n \to \infty$.
Let
  \begin{equation}
  s_n=\log r_n, f^{(n)}(r)=1 + e^{2s} \phi^{(n)}(s), 
  \rho^{(n)}(r) = e^{s} \psi^{(n)}(s).
  \end{equation}
Then $\phi^{(n)}, \psi^{(n)}$ have uniform bounds in $(-\infty, s_n]$.
But by  (\ref{E:g11}), we have $\psi^{(n)}(s_n) \to + \infty$, 
as $n \to \infty$,
which is a contradiction. So $\bar{r} > 0$.

   Now choose $0 <\hat{r} < \infty$ so that (\ref{E:g50}) is satisfied for
$r \in [0, \hat{r} ]$ for all $n$. Then still by  (\ref{E:g11})  we have that
$\psi^{(n)}(s) \to + \infty$, as $n \to \infty$, for all
$s \in ( - \infty, \log(\hat{r}) ]$. This contradiction implies that 
the lemma is proved. 
\setcounter{equation}{0}
\section*{8. Proof of the Theorem }
  
 By Lemma 9   there is $\beta_1 >0$ so that if
 $\alpha \notin S_{\beta_1} ^{-} \cup S_{\beta_1}^{+}$ , (A) is satisfied. 
By Lemma 10 there is $\beta_2 > \beta_1 >0$ so that if 
$\alpha \notin S_{\beta_2} ^{-} \cup S_{\beta_2}^{+}$ , (B) is satisfied. 
For $\beta \in [\beta_1, \beta_2]$, by Lemma 2,3, there are 
$0 < \alpha_1 < \alpha_2$,
so that if $\alpha \in (0, \alpha_1]$, then $\alpha \in S_\beta^{-}$, and
if   $\alpha \in [ \alpha_2, \infty)$, then $\alpha \in S_\beta^{+}$.
Define 
   \begin{equation}
      I = [\alpha_1, \alpha_2] \times  [\beta_1, \beta_2].
   \end{equation}
Let
   \begin{eqnarray}
   & &  S^{-} = \{ (\alpha, \beta) \in I |
        \textnormal{ $f'$ cross 0 before $f$ cross 0} \},  \\ 
   & &  S^{+} = \{ (\alpha, \beta) \in I |
       \textnormal{ $f$ cross 0 before $f'$ cross 0} \}.
   \end{eqnarray}
We have seen that $S^{-}, S^{+}$ are open, nonempty and disjoint. 
 
  By the topological lemma in \cite{mcleod}(McLeod and Serrin) , 
there is a continuum $\Gamma$
connects $\beta=\beta_1$ and $\beta=\beta_2$. Define
  \begin{eqnarray}
  & & \Omega_-= \{ (\alpha,\beta) \in \Gamma |
   \textnormal{ (A) is satisfied} \},  \\
  & & \Omega_+= \{ (\alpha,\beta) \in \Gamma |
   \textnormal{ (B) is satisfied} \}.
  \end{eqnarray}
By the choice of $\beta_1,\beta_2$,   we see that 
$ \Omega_- ,  \Omega_+ $ are not empty and disjoint,
and it's easy to show that  $ \Omega_- ,  \Omega_+ $  are
open in $\Gamma$. Thus there exists $(\alpha^{*}, \beta^{*}) \in 
\Gamma \setminus ( \Omega_- \cup  \Omega_+ $, such that (C) is satisfied.
So $f(r, \alpha^*, \beta^*), \rho(r, \alpha^*, \beta^*)$ is a solution 
to the boundary value problem, and satisfying the properties   
  \begin{eqnarray}
     0<f<1, f' \le 0,  \\
     0<\rho<\rho_0, \rho' \ge 0,
  \end{eqnarray}
for $0<r<\infty$.
 
   Finally we need to show $f' <0, \rho'>0$.

   Suppose $r_1$ is a positive zero of $f'$. Then by eq. (1.10),
there is
 $$f'''(r_1) = 2 \left( {1-f^2 \over r^3} + g_0^2 \rho \rho' \right) f >0,$$
which is a contradiction.  So $f'<0$.

  Now suppose $r_2$ is a zero point of $\rho'$. Then because
$\rho' \ge 0$, there is $\rho''(r_2) =0$. By eq. (1.11), we have
  $$ \rho'''(r_2)=4 \left( {f f' \over r^2} -
            {f^2 \over r^3}  \right) \rho < 0,  $$
which is a contradiction.
So the theorem is proved.


\end{document}